\documentclass{elsart}
\usepackage{epsfig,graphicx,amsfonts,amssymb}
\begin{document}
\begin{frontmatter}
\title{Stock market return distributions: from past to present}

\author{S.~Dro\.zd\.z$^{1,2}$, M.~Forczek$^{1}$, J.~Kwapie\'n$^1$, 
P.~O\'swi\c ecimka$^{1}$, R.~Rak$^{2}$}

\address{$^1$Institute of Nuclear Physics, Polish Academy of Sciences,
PL--31-342 Krak\'ow, Poland}
\address{$^2$ Institute of Physics, University of Rzesz\'ow, PL--35-959 
Rzesz\'ow, Poland}

\begin{abstract}

We show that recent stock market fluctuations are characterized by the 
cumulative distributions whose tails on short, minute time scales exhibit 
power scaling with the scaling index $\alpha > 3$ and this index tends to 
increase quickly with decreasing sampling frequency. Our study is based on 
high-frequency recordings of the S\&P500, DAX and WIG20 indices over the 
interval May 2004 - May 2006. Our findings suggest that dynamics of the 
contemporary market may differ from the one observed in the past. This 
effect indicates a constantly increasing efficiency of world markets. 

\end{abstract}


\end{frontmatter}

The so-called financial stylized facts are among the central issues of 
econophysics research. Much effort has been devoted on both the empirical 
and the theoretical level to such phenomena like fat-tailed distributions 
of financial fluctuations, persistent correlations in volatility, 
multifractal properties of returns etc. Specifically, the interest in the 
return distributions can be traced back to an early work of 
Mandelbrot~\cite{mandelbrot63} in which he proposed a L\'evy process as 
the one governing the logarithmic price fluctuations. Much later this 
issue was revisited in~\cite{mantegna95} based on data with much better 
statistics and a new model of exponentially-truncated L\'evy flights was 
introduced. Then, in an extensive systematic study of the largest American 
stock markets~\cite{gopi99} the distribution tails for both the prices and 
the indices were shown to be power-law with the scaling exponent $\alpha 
\simeq 3$. The most striking outcome of that study was that despite the 
fact that the tails were well outside the L\'evy-stable regime ($\alpha 
\le 2$), they were apparently stable under time aggregation up to several 
days for indices and up to a month for stocks. The existence of return 
distributions with scaling tails was also reported in other markets like 
e.g. London~\cite{farmer04}, Frankfurt~\cite{lux96}, 
Paris~\cite{gabaix03}, Oslo~\cite{skjeltorp00}, Tokyo~\cite{gopi99}, and 
Hong Kong~\cite{gopi99,huang00} but sometimes with a slightly different 
value of the scaling index. This empirical property of price and index 
returns led to the formulation of the so-called inverse cubic power 
law~\cite{gopi99}, which was soon followed by an attempt of formulating 
its theoretical foundation~\cite{gabaix03} (see also~\cite{farmer04}).

Subsequent related study~\cite{drozdz03} revealed that, opposite to the 
earlier outcomes of~\cite{gopi99}, the tail shape of the return 
distributions might no longer be so stable along time axis. After 
comparison of the results obtained from the American stock market data in 
years 1994-95 and in 1998-99, it turned out that in more recent data the 
scaling tails with $\alpha \simeq 3$ for individual companies are 
preserved up to the time scales (sampling intervals) $\Delta t$ of less 
than one hour instead of one month. This earlier crossover for 1998-99 
data can easily be seen in Fig.~1. This result was obtained by extending 
our previous analysis~\cite{drozdz03} over a set of 1000 largest 
American companies~\cite{taq} in order to enable more direct comparison 
with outcomes of ref.~\cite{gopi99} based on the same number of stocks. 
The inverse cubic scaling is still evident in Fig.~1 for short time 
scales up to $\Delta t=4$ min, but the scaling index starts rising already 
for data with $\Delta t = 16$ min and for longer time scales the tail 
behaviour is clearly governed by the Central Limit Theorem. The difference 
of the results for the same market but for different time intervals might 
suggest that the scaling behaviour is not stable and depends on some 
crucial factors as, for example, the speed of information processing which 
constantly increases from past to present.

\begin{figure}
\hspace{1.0cm}
\epsfxsize 12.5cm
\epsffile{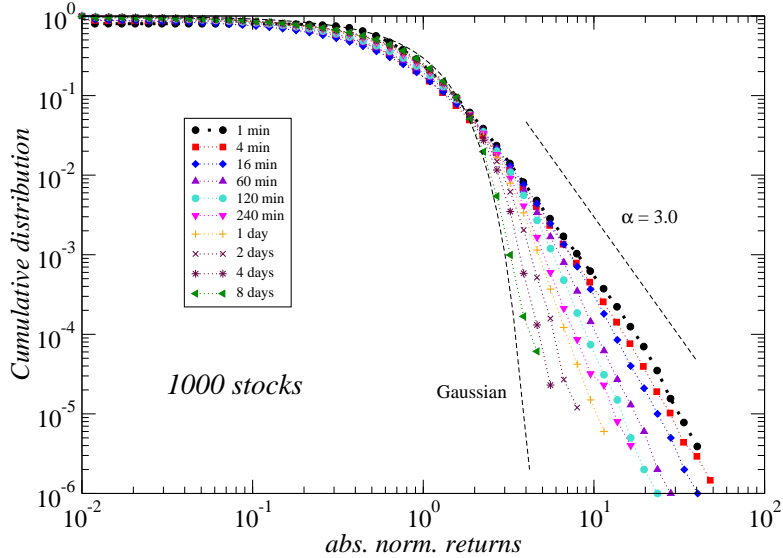}
\caption{Cumulative distributions of normalized stock returns averaged 
over 1000 highly-capitalized American companies in time interval Dec 1997 
- Dec 1999 for several different time scales from 1 min to 8 days. 
Gaussian distribution and inverse cubic scaling are also shown for 
comparison. Best-fit power index $\alpha$ calculated by means of log-log regression 
assumes the following values: $3.08 \pm 0.05$ ($\Delta t = 1$ min), $3.34 
\pm 0.05$ (4 min), $4.00 \pm 0.04$ (16 min), $4.60 \pm 0.05$ (60 min), 
$4.95 \pm 0.06$ (120 min), $4.81 \pm 0.15$ (240 min), $5.90 \pm 0.08$ (1 
day), $7.18 \pm 0.28$ (2 days), $9.17 \pm 0.22$ (4 days), and $8.32 \pm 
0.40$ (8 days).} 
\end{figure}

This possibility can further be examined by considering even more recent 
data from the American market. First, we look at the S\&P500 index which 
already was the subject of an analysis in~\cite{gopi99}. Our data is a 
time series of 1 min returns covering the period May 2004 - May 2006 
(in~\cite{gopi99} the period was 1984-1996). The c.d.f. for this data is 
presented in Fig.~2(a) for several time scales up to 120 min. The most 
interesting feature is the lack of inverse cubic scaling even for the 
shortest one-minute returns: in this case the actual scaling index is 
slightly above 4 and it systematically increases with decreasing sampling 
frequency (see Table 1). We leave open here the question what factor 
underlies the evident absence of the $\alpha \simeq 3$ type of scaling: 
the dynamics of S\&P500 returns could have changed sufficiently 
significantly since earlier half of 1990s and the inverse cubic scaling no 
longer exists or it still exists but is restricted to time scales shorter 
than 1 minute. That our observation is more universal and can be made for 
other markets as well one may infer from Fig.~2(b) and Fig.~2(c) 
presenting c.d.f. for the German index DAX and for the Polish index WIG20, 
respectively, for the same period of time. While the returns of DAX did 
not exactly comply with the inverse cubic scaling also in the period 
1998-99~\cite{drozdz03}, the ones of WIG20 indeed used to display this 
kind of behaviour in the past as documented in~\cite{rak07}. However, 
nowadays WIG20 also develops much thinner tails with $\alpha > 4$ 
for $\Delta t=1$ min (Table 1). 

\begin{table}
\begin{tabular}{|c||c|c|c|c|c|}
\hline
$\Delta t$ & 1 min & 4 min & 16 min & 32 min & 60 min \\
\hline\hline
S\&P500 & $4.12 \pm 0.12$ & $4.21 \pm 0.08$ & $5.18 \pm 0.21$ & $5.53 \pm 0.20$ & $6.10 
\pm 0.35$ \\
\hline
DAX & $3.56 \pm 0.12$ & $3.76 \pm 0.05$ & $4.44 \pm 0.16$ & $5.14 \pm 0.26$ & $5.16 \pm 
0.66$ \\
\hline
WIG20 & $4.28 \pm 0.16$ & $5.24 \pm 0.27$ & $5.81 \pm 0.78$ & $5.61 \pm 0.45$ & $6.30 \pm 
0.72$ \\
\hline\hline
\end{tabular}
\caption{Values of the scaling index $\alpha$ obtained with a log-log regression fit for
three different market indices (S\&P500, DAX, and WIG20) from the interval May 2004 - May 
2006.}
\end{table}

\begin{figure}
\hspace{-0.5cm}
\epsfxsize 6.5cm
\epsffile{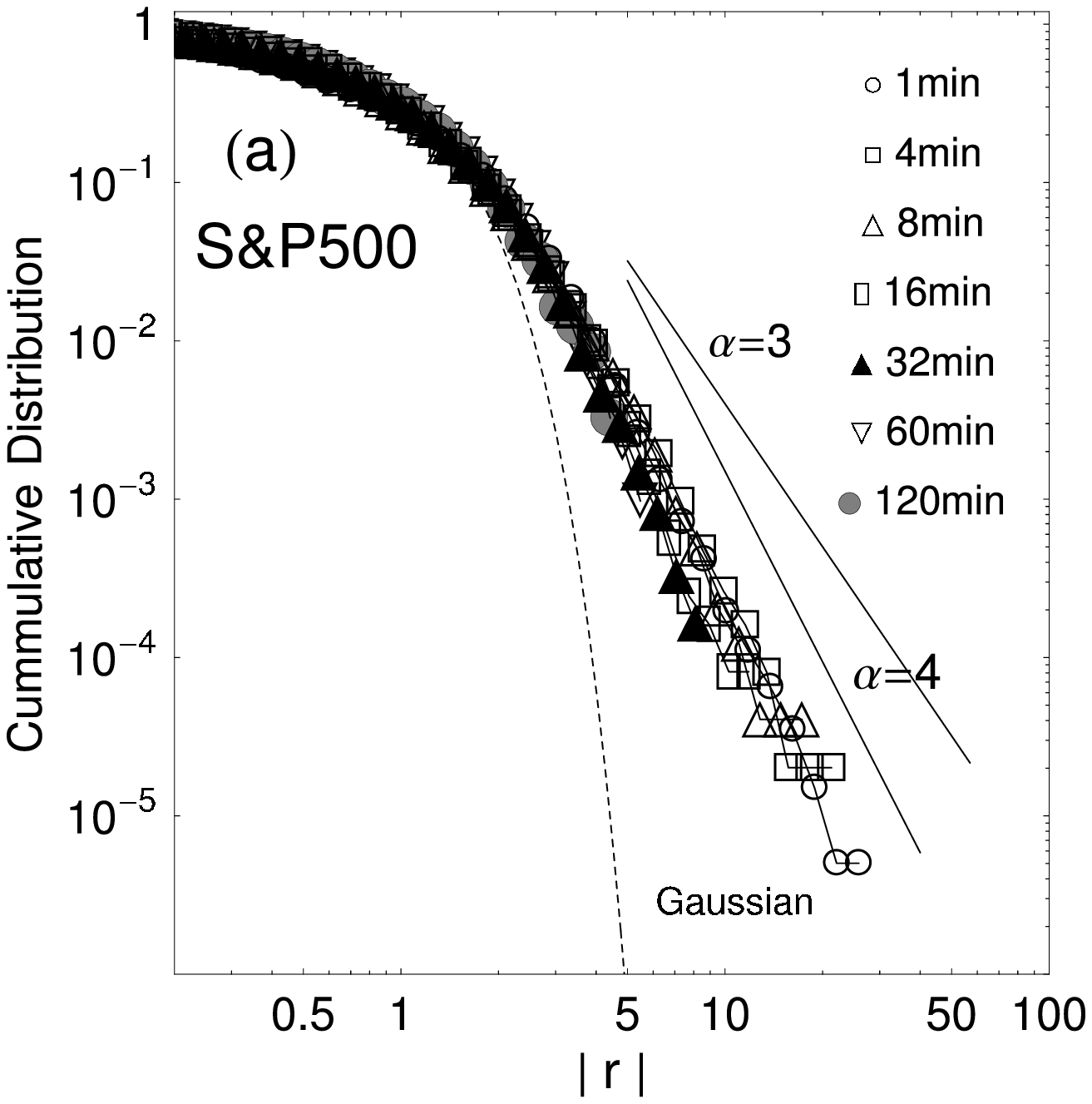}
\epsfxsize 6.5cm
\epsffile{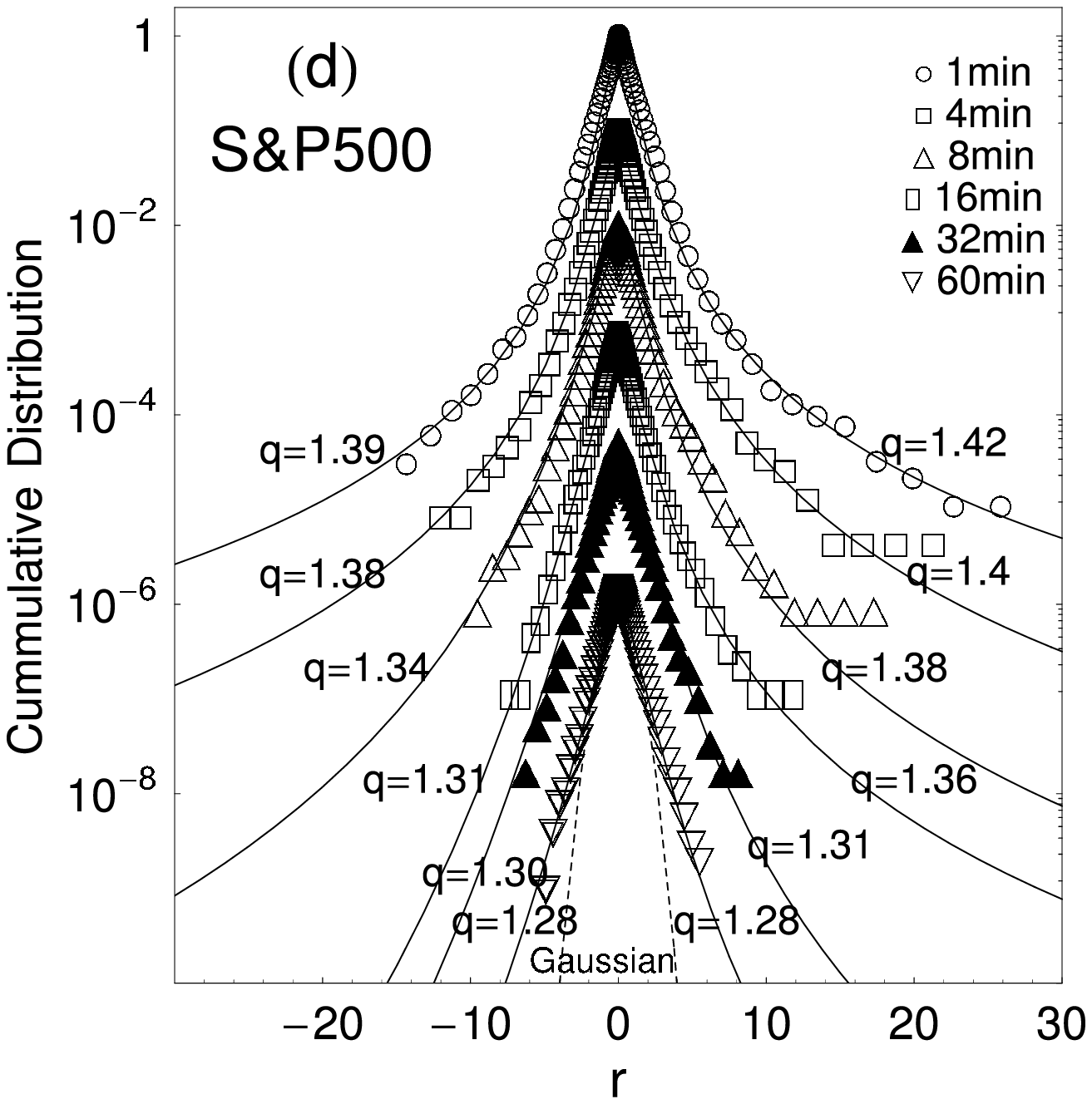}

\hspace{-0.5cm}
\epsfxsize 6.5cm
\epsffile{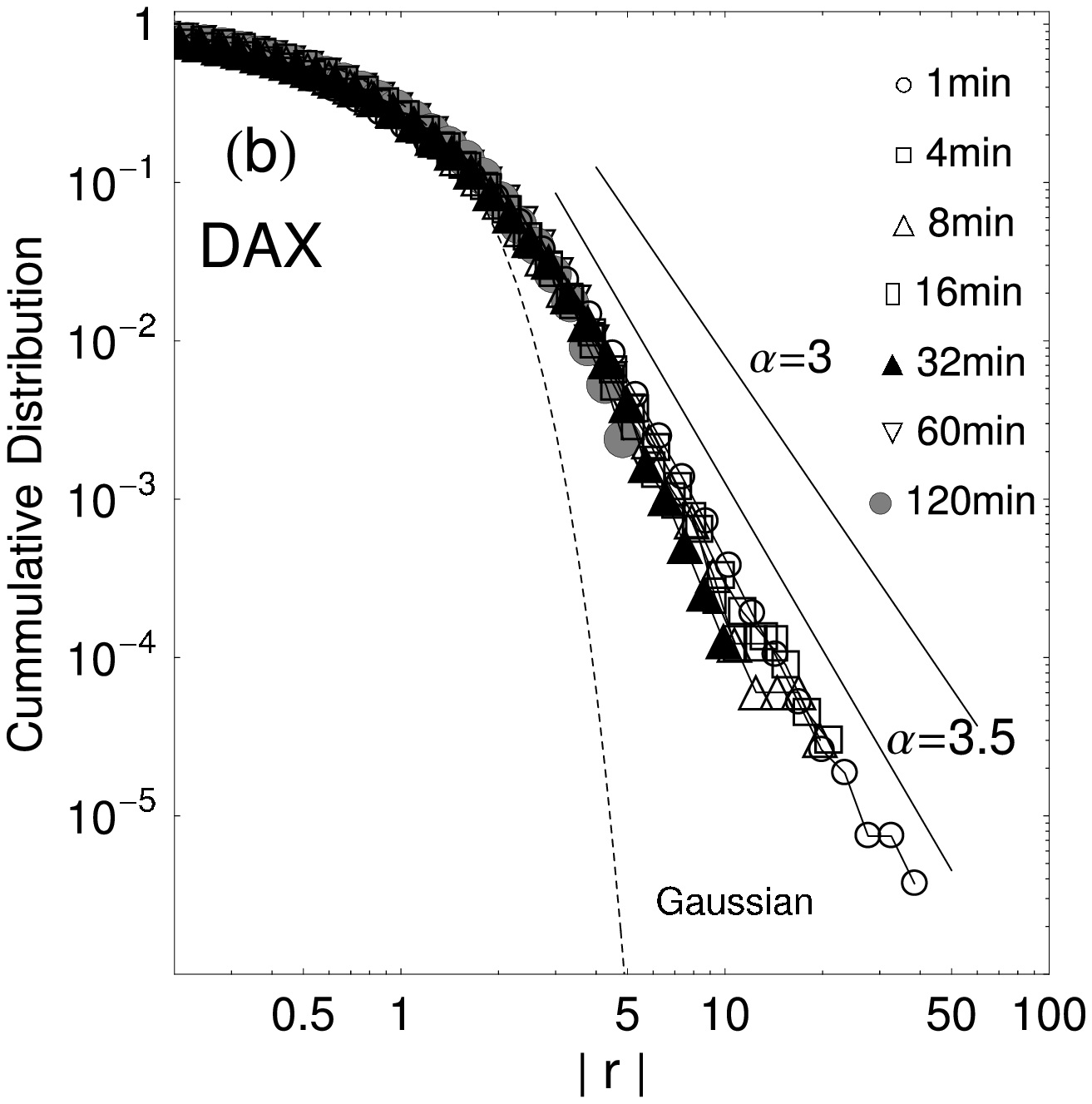}
\epsfxsize 6.5cm
\epsffile{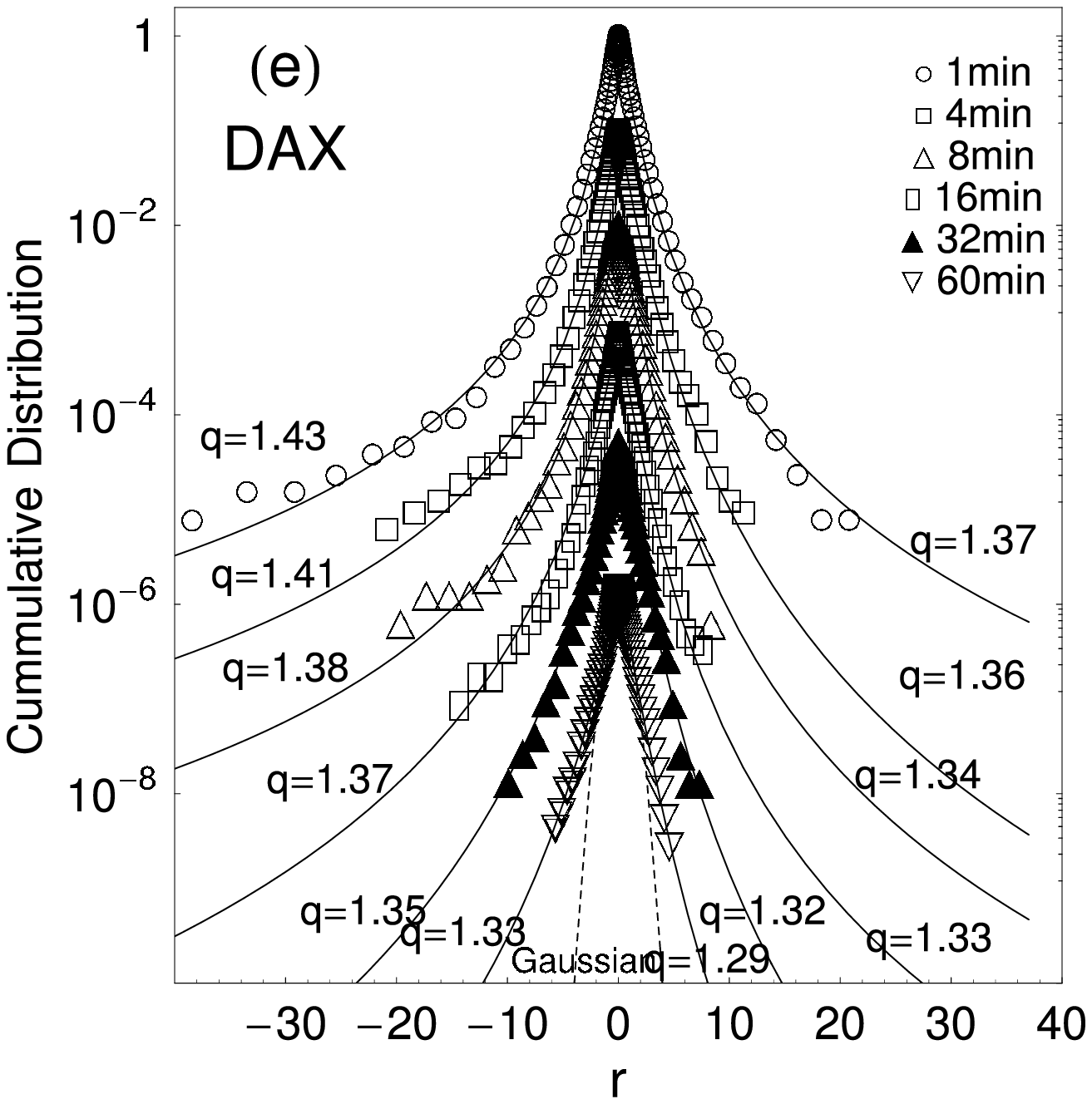}

\hspace{-0.5cm}
\epsfxsize 6.5cm
\epsffile{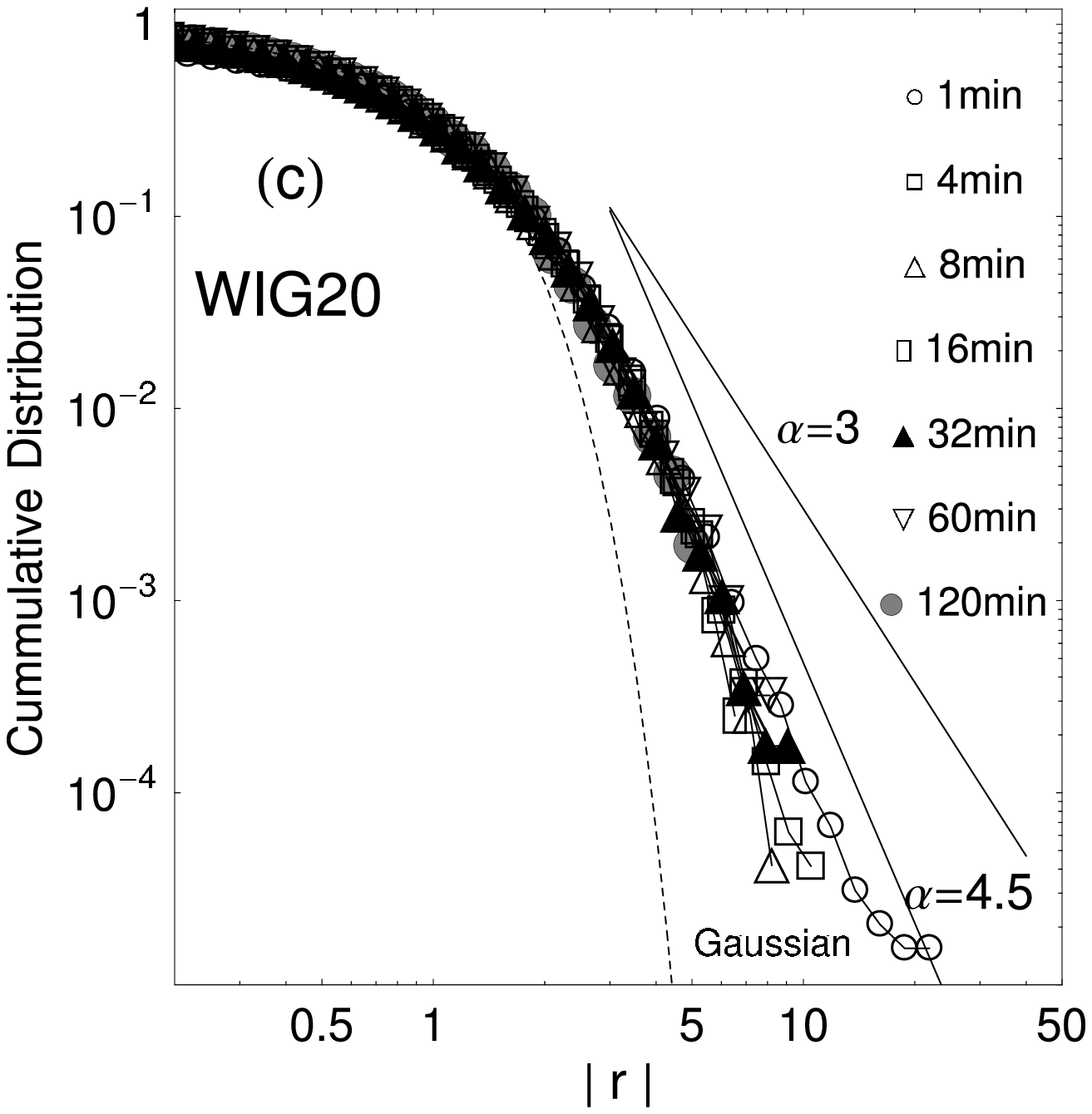}
\epsfxsize 6.5cm
\epsffile{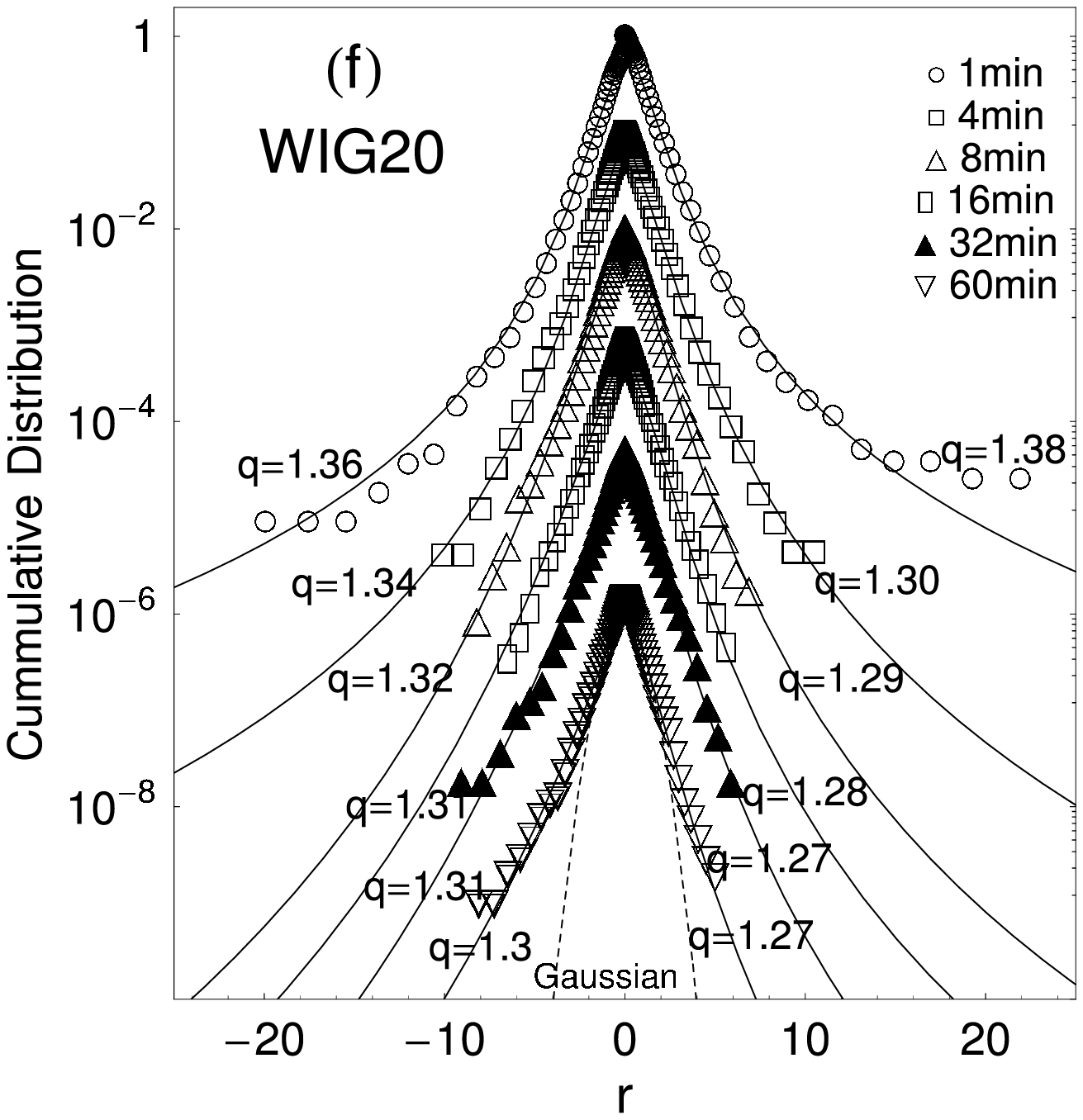}
\caption{Cumulative distributions of index returns for American S\&P500 
(top row), German DAX (middle row) and Polish WIG20 (bottom row) recorded  
over the interval May 2004 - May 2006. Data points for sampling intervals 
from 1 min to 120 min are denoted by different symbols. (Left column) 
Distributions for normalized returns are shown together with Gaussian 
distribution and lines corresponding to inverse cubic scaling and, 
approximately, the actual scaling. (Right column) Experimental 
distributions are best-fitted by $q$-Gaussians with a free parameter $q$ 
(fitted values displayed) separately for negative and positive returns. 
Note the small asymmetry between left and right tails.}
\end{figure}

Recent studies~\cite{rak07,tsallis03} showed that in a wide range of 
returns the financial return distributions can be approximated by a family 
of $q$-Gaussians~\cite{tsallis98} with the parameter $q$ depending on the 
sampling interval $\Delta t$. The $q$-Gaussian distributions follow 
naturally from the nonextensive statistical 
mechanics~\cite{tsallis98,tsallis03} and their c.d.f. can be written 
as~\cite{rak07}
\begin{equation}
P(X > x) = \mathcal{N}_q \left({\sqrt{\pi}\Gamma({1 \over 2} (3 - 
q)\beta) \over 2 \Gamma(\beta)\sqrt{\mathcal{B}_q \over \beta)}} \pm (x - 
\bar{\mu_q})_2 F_1 (\alpha,\beta;\gamma;\delta) \right),
\label{cdf}
\end{equation}
where $\alpha={1 \over 2}$, $\beta=1/(q-1)$, $\gamma = {3 \over 2}$, 
$\delta = -\mathcal{B}_q(q-1)(\bar{\mu}_q - x)^2$, ${}_2 F_1 
(\alpha,\beta;\gamma;\delta) = \sum_{k=0}^{\infty} 
{\delta^k(\alpha)_k(\beta)_k \over k!(\gamma)_k}$ is the Gauss 
hypergeometric function and $\mathcal{N}_q$, $\bar{\mu}_q$ are, 
respectively, the normalization factor and $q$-mean of the $q$-Gaussian 
p.d.f.~\cite{tsallis98}. Fig.~2(d)-2(f) exhibit cumulative distributions 
of returns for the same three indices with the corresponding best fits in 
terms of Eq.(\ref{cdf}). The theoretical curves are in satisfactory 
agreement with the data for all the considered values of $\Delta t$. It 
is noteworthy that, consistently with the left-hand side panels of 
Fig.~2, the largest values of $q$ are well below $q = 3/2$, which 
corresponds to the inverse cubic scaling, and decrease with decreasing 
sampling frequency towards the classic Gaussian distribution with $q = 
1$.

\begin{figure}
\hspace{2.0cm}
\epsfxsize 8.0cm
\epsffile{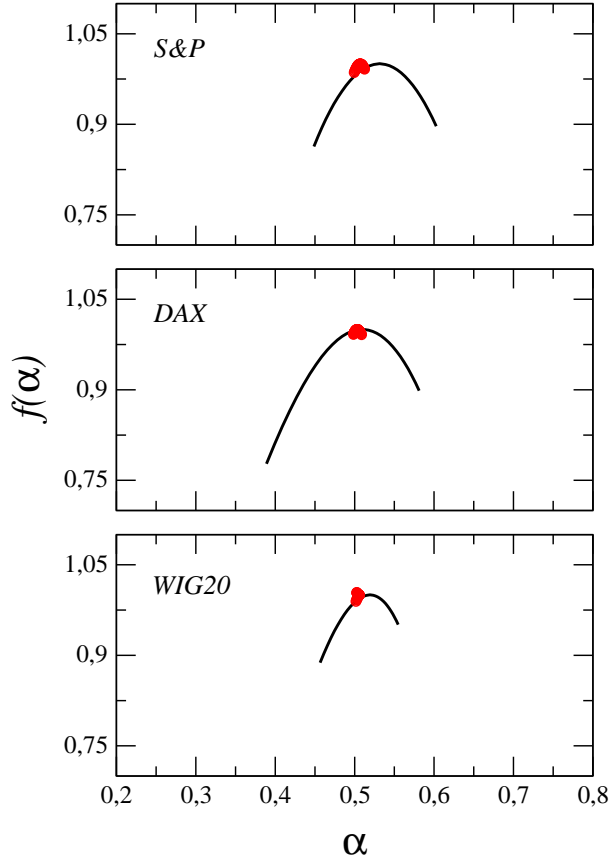}
\caption{Singularity spectra for 1 min index returns: S\&P500 (top), DAX 
(middle), and WIG20 (bottom). In each case, the actual data (solid line) 
is accompanied by its randomized version averaged over 10 independent 
realizations (full symbols).}
\end{figure}

Finally, we look at the singularity spectra $f(\alpha)$ (for numerical 
details of the method see e.g.~\cite{kwapien05}) of our time series under 
study. Two cases are considered: original data comprising the full 
variety of nonlinear temporal correlations (solid lines in Fig.~3) and 
the randomized data in which all the correlations are removed by 
shuffling the data points (full symbols in Fig.~3). Since both the 
nonlinear dependencies and the fat-tailed distributions can be potential 
sources of multifractality, the data shuffling can give some information 
of how rich is the multiscaling behaviour due to each of these 
sources~\cite{matia03,kwapien05}. In the present context the most 
interesting feature is that the Gaussian distribution of uncorrelated 
data is associated with a monofractal $f(\alpha)$ spectrum. Fig.~3 shows 
the singularity spectra for S\&P500, DAX and WIG20 (top to bottom) 
together with their randomized-data counterparts. In all three cases, the 
original data clearly represent multifractal processes with DAX and 
S\&P500 showing richer multifractality than WIG20. This picture changes 
completely if we look at the randomized data: we cannot detect 
sufficiently significant trace of multifractality ($f(\alpha)$ is almost 
point-like). This is in agreement with the observation that the 
distribution tails for the contemporary data tend to be thinner than  
before. This result gives also an additional argument in favour of the 
statement that the principal (here even the unique) source of 
multifractal properties of the stock market data are the nonlinear 
correlations.

In this paper we have shown that c.d.f. for the most recent stock market 
data represented by the index returns develops tails whose scaling index 
rises above the value of 3 even for short, minute sampling intervals. This 
means that contemporary market dynamics significantly differs from the one 
observed 20 or even 10 years ago and described by the inverse cubic power 
law~\cite{gopi99}. That these changes are a continuous process rather that 
a sudden transition we infer from the existence of intermediate stages in 
which the inverse cubic scaling was observed up to medium time scales of 
tens of minutes but was absent in daily data~\cite{drozdz03,rak07}. This 
effect suggests a scenario of constantly increasing market efficiency due 
to an acceleration of information processing in the world 
markets~\cite{drozdz03,kwapien04}.  The related compression of the range of 
potential time correlations between consecutive returns in the present 
analysis finds evidence in a faster convergence towards a Gaussian 
distribution for aggregated returns. Such a faster convergence indicates 
weeker time-correlations between returns. Further argument in favour of an 
increasing stock market efficiency comes from the autocorrelation 
analysis. The autocorrelation functions calculated exlicitely from the 1 
min returns for the three indices considered above drop down to the noise 
level already for time-lags as small as 1-2 min. This is to be compared to 
$\sim 5$ min in the period 1998-99~\cite{drozdz03}, and to $\sim 20$ min 
which according to ref.~\cite{gopi99} was characteristic for the period 
1994-95.


\begin{thebibliography}{}

\bibitem{mandelbrot63} B.~Mandelbrot, J.~Business {\bf 36}, 294 (1963)

\bibitem{mantegna95} R.N.~Mantegna, H.E.~Stanley, Nature {\bf 376} (1995) 46

\bibitem{gopi99} P.~Gopikrishnan, V.~Plerou, L.A.~Nunes Amaral, M.~Meyer, 
H.E.~Stanley, Phys. Rev. E {\bf 60} (1999) 5305; V.~Plerou, 
P.~Gopikrishnan, L.A.N.~Amaral, M.~Meyer, H.E.~Stanley, Phys. Rev. E {\bf 
60} (1999) 6519

\bibitem{farmer04} J.D.~Farmer, F.~Lillo, Quant.~Finance {\bf 4} (2004) 
C7; J.D.~Farmer, L.~Gillemot, F.~Lillo, S.~Mike, A.~Sen, Quant.~Finance 
{\bf 4} (2004) 383

\bibitem{lux96} T.~Lux, Appl. Financial Economics {\bf 6} (1996) 463

\bibitem{gabaix03} X.~Gabaix, P.~Gopikrishnan, V.~Plerou, H.E.~Stanley, 
Nature {\bf 423} (2003) 267

\bibitem{skjeltorp00} J.A.~Skjeltorp, Physica A {\bf 283} (2000) 486

\bibitem{huang00} Z.F.~Huang, Physica A {\bf 287} (2000) 405

\bibitem{drozdz03} S.~Dro\.zd\.z, J.~Kwapie\'n, F.~Gr\"ummer, F.~Ruf, 
J.~Speth, Acta Phys. Pol. B {\bf 34} (2003) 4293, cond-mat/0208240

\bibitem{taq} See http://www.taq.com

\bibitem{rak07} R.~Rak, S.~Dro\.zd\.z, J.~Kwapie\'n, Physica A {\bf 374} 
(2007) 315

\bibitem{tsallis03} C.~Tsallis, C.~Anteneodo, L.~Borland, R.~Osorio, 
Physica A {\bf 324} (2003) 89

\bibitem{tsallis98} C.~Tsallis, R.S.~Mendes, A.R.~Plastino, Physica A 
{\bf 261} (1998) 534

\bibitem{matia03} K.~Matia, Y.~Ashkenazy, H.E.~Stanley, Europhys.~Lett. 
{\bf 61} (2003) 422

\bibitem{kwapien05} J.~Kwapie\'n, P.~O\'swi\c ecimka, S.~Dro\.zd\.z, 
Physica A {\bf 350} (2005) 466

\bibitem{kwapien04} J.~Kwapie\'n, S.~Dro\.zd\.z, J.~Speth, Physica A {\bf 
337} (2004) 231

\end{thebibliography}
\end{document}